\documentclass[aps,prb,twocolumn,superscriptaddress,showpacs]{revtex4-1}

\usepackage{amsmath}
\usepackage{graphicx}
\usepackage{color}
\usepackage{hyperref}

\begin{document}

\title{Electron-phonon coupling in metallic carbon nanotubes: \\
Dispersionless electron propagation despite dissipation}

\author{Roberto Rosati}
\affiliation{
Department of Applied Science and Technology, Politecnico di Torino \\
C.so Duca degli Abruzzi 24, 10129 Torino, Italy
}

\author{Fabrizio Dolcini}
\affiliation{
Department of Applied Science and Technology, Politecnico di Torino \\
C.so Duca degli Abruzzi 24, 10129 Torino, Italy
}
\affiliation{CNR-SPIN, Monte S.Angelo - via Cinthia, I-80126 Napoli, Italy}

\author{Fausto Rossi}
\affiliation{
Department of Applied Science and Technology, Politecnico di Torino \\
C.so Duca degli Abruzzi 24, 10129 Torino, Italy
}

\date{\today}

\begin{abstract}
A recent study [Rosati, Dolcini, and Rossi, Appl. Phys. Lett. {\bf 106}, 243101 (2015)] has predicted that, while in semiconducting single-walled carbon nanotubes (SWNTs) an electronic wave packet experiences the typical spatial diffusion of conventional materials, in metallic SWNTs its shape remains essentially unaltered up to micrometer distances at room temperature, even in the presence of the electron-phonon coupling. Here, by utilizing a Lindblad-based density-matrix approach  enabling us to account for both  dissipation and decoherence effects, we test such a prediction by analyzing various aspects that were so far unexplored. In particular, accounting for initial nonequilibrium excitations, characterized by an excess energy $E_0$, and including both intra- and interband phonon scattering, we show that for realistically high values of $E_0$ the electronic diffusion is extremely small and nearly independent of its energetic distribution, in spite of a significant energy-dissipation and decoherence dynamics. Furthermore, we demonstrate that the effect is robust with respect to the variation of the chemical potential. Our results thus suggest that metallic SWNTs are a promising platform to realise quantum channels for the non-dispersive transmission of electronic wave packets.
\end{abstract}



\maketitle

\section{Introduction}\label{s-I}

Using wave dynamics as a platform to encode  information naturally offers the possibility to exploit the superposition of states and, thereby, to perform an intrinsically parallel transfer and manipulation of information. To this purpose, a crucial ingredient is to generate sequences of wavepackets  propagating coherently without overlapping to each other. In quantum optics, where sources of single-photon wave packets  have been achieved since long, the control of light propagation and polarization with beam splitters and polarizers is extremely high, and photonic materials are nowadays considered a realistic platform  to perform  scalable quantum computing\cite{Kok07a}. \\
The exciting perspective to achieve a similar degree of control using electron waves\cite{Fève07a,McNeil11a} has led to the implementation of   single electron pumps  with various setups\cite{Kouwenhoven91a,Pothier92a,Shilton96a,Pekola08a}. However, despite a number of proposals\cite{Bertoni00a,Ionicioiu01b,Jefferson06a,Fève08a,Pekola13a,Ott14a} the realization of flying qubits via single-electron wave packets of controllable shape and phase that propagate ballistically in low-dimensional conductors   still remains a fascinating  challenge in Physics.

A major difference between an electromagnetic and an electronic wave  is that, while   the velocity of a photon is nearly independent of its wavevector $\mathbf{k}$,  the group velocity of an electron in conventional materials --characterized by a paraboliclike dispersion relation-- depends on $\mathbf{k}$, so that its components propagate with different velocities. This leads to an intrinsic spreading of an electron wavepacket, even in the absence of scattering processes.
However, in metallic single-walled carbon nanotubes (SWNTs), in graphene, and in the surface states of topological insulators, electrons behave as massless relativistic fermions and, just like photons, are characterized by a linear spectrum, with the Fermi velocity $v_F \sim 10^6\,{\rm m/s}$ playing the role of the speed of light $c$. This property makes such materials ideal candidates for an electronic alternative to photon-based quantum information processing.
In graphene, for instance, electron supercollimation has been predicted to occur when an external static and long-range disorder 
is suitably applied.~\cite{Park08a,Choi14a} 
SWNTs are even more promising, in view of the accuracy reached in their synthesis,\cite{Maruyama02a,Arnold06a} their behavior as one-dimensional ballistic conductors,\cite{Tans97a,Javey03a} and their versatility in forming perfectly aligned arrays for high-performance electronic devices.\cite{Chen06a,Kang07a,Nougaret09a} 

Although in principle an electron wave packet can propagate along a metallic SWNT maintaining its initial shape, in realistic devices   such property may be affected by scattering processes. Extrinsic scattering due to impurities can nowadays be made essentially negligible, by exploiting well established  fabrication techniques yielding ultra-clean nanotubes by avoiding exposure to chemicals.\cite{Maruyama02a,Laird15a} Intrinsic scattering mechanisms involve electron-electron   and electron-phonon couplings. The former plays an important role at very low temperatures, where it has been shown to lead to the Coulomb blockade\cite{Tans98a} and Luttinger liquid behavior\cite{Bockrath99a} .
At intermediate and room temperature, however, electron-phonon coupling is the most important scattering mechanism, as experimental results indicate.\cite{Yao00a,Park04a,Javey04a} 
For these reasons, in the last few years various theoretical studies have analyzed the effects of electron-phonon coupling in SWNTs. 
On the one hand, models based on a classical-like treatment of the electron-phonon coupling as an external oscillating potential\cite{Roche05a,Roche05b,Ishii07a,Ishii08a} enable one to analyze the time-dependent evolution of  single wave packets and to obtain  the linear conductance by performing a suitable averaging over the initial state. These approaches, however, fail in capturing the intrinsically dissipative nature of the phonon bath. On the other hand, treating electron-phonon coupling in SWNTs via the Boltzmann-equation  schemes\cite{Bonilla14a,Perebeinos05a} does not allow one to account for electronic phase coherence.   

In a recent work\cite{Rosati15a}, it has been shown that, while in semiconducting SWNTs an electronic wavepacket  spreads already for a scattering-free propagation, in  metallic SWNTs the shape of the wavepacket can remain essentially unaltered, even in the presence of  electron-phonon coupling, up to micrometer distances at room temperature.
Although such a result is quite promising, a number of fundamental questions remain still open in the problem. In the first instance, the case of nonequilibrium carrier distributions has not been discussed so far. Secondly, the result of Ref.[\onlinecite{Rosati15a}] is limited to the case of intraband phonon scattering, whereas interband coupling may be significant, especially due to breathing phonon modes.   Furthermore, while the spatial dynamics of the wavepacket has been discussed, it is still unclear how  dissipation and decoherence affect its energy and momentum distribution. Finally, it is crucial to understand   whether and to what extent the predicted dispersionless propagation is affected by a change of the chemical potential.  

This paper addresses these relevant problems. 
To this purpose, we apply a recently developed density-matrix approach\cite{Taj09b,Rosati14e} that enables us to account for  both energy-dissipation and  decoherence effects. Focussing on the case of a metallic SWNT, we demonstrate that the shape of the wavepacket is essentially unaltered, even in the presence of   interband electron-phonon coupling, provided that the excess energy of the excitation is realistically high. Thus, despite a significant energy-dissipation and decoherence dynamics, the electronic diffusion in metallic SWNT is extremely small and nearly independent of the wavepacket energetic distribution. Furthermore, we show that this effect is weakly dependent on the chemical potential, at least at room temperature. Our results thus support the conclusion  that metallic SWNTs can be considered as an electron-based platform for information transfer.

The paper is organized as follows: In Sec.~\ref{s-PS}, we  describe the SWNT model utilized to account for the electronic and phononic energy spectrum, as well as for the corresponding electron-phonon coupling. In Sec.~\ref{s-TA} we   briefly summarize the main aspects of the Lindblad-based density-matrix formalism developed in Ref.[\onlinecite{Rosati14e}], providing the explicit expression for the electronic properties needed for the present investigation, namely the spatial and the energetic carrier distributions.
In Sec.~\ref{s-SE}, we   present simulated experiments that enables us to quantify the impact of intra- as well as interband carrier-phonon interactions on the propagation of electron wavepackets for different initial conditions and chemical-potential values. As we shall discuss, the highly nontrivial interplay between energy dissipation and electronic quantum diffusion is crucial for such a purpose.
Finally, in Sec.~\ref{s-SC}, we  summarize our results and draw the conclusions.

\section{SWNT model}\label{s-PS}

In order to describe our SWNT, we adopt the well established model developed by Ando and co-workers (see Ref. [\onlinecite{Ando05a}] and references therein), whose  main ingredients needed for our analysis are summarized here below.

{\it Electronic properties}. The low-energy electron dynamics in a SWNT decouples into two valleys around the $\mathbf{K}$ and $\mathbf{K}^\prime$ points, described by the following Hamiltonian matrices in the sublattice basis, 
\begin{equation}\label{H1}
H_\mathbf{K}=\hbar v_F \boldsymbol\sigma \cdot  \mathbf{k}\ , \qquad
H_{\mathbf{K}^\prime}= -\hbar v_F \boldsymbol\sigma^* \cdot  \mathbf{k}\ ,
\end{equation} 
where  $\boldsymbol\sigma =(\sigma_x,\sigma_y)$ denote Pauli matrices acting on the twofold sublattice space, and  $\mathbf{k}=(k_{n,\nu}^\perp,k)$  the  carrier wavevector.\cite{Ando05a} Here $k$ denotes the continuous component along the SWNT axis  ($\parallel$), whereas $k_{n,\nu}^\perp=(n+v \nu/3)/R$ is the discrete component along the circumference ($\perp$), where~$n$  is the electron subband, $v =\pm 1$ for  the $\mathbf{K}/\mathbf{K}^\prime$ valley, $R$ the nanotube radius, and the index $\nu = 0, \pm 1$ is defined through the relation $\exp[i \mathbf{K} \cdot \mathbf{C}]=\exp[-i \mathbf{K}' \cdot \mathbf{C}]=\exp[2\pi i \nu/3]$, where $\mathbf{C}$ is the vector rolling    the graphene lattice into the SWNT.

Since typical subband energy separations are of the order of ${\rm eV}$, we shall focus on the lowest energy subband ($n=0$), whose energy spectrum is independent of the valley $v={\bf K/K^\prime}=\pm1$ and is given by
\begin{equation}\label{epsilon}
\epsilon_\alpha = b \,\hbar v_F \sqrt{
k^2 +( \nu/3R)^2
}\ ,
\end{equation}
where $\alpha=(k, b)$ is the quantum number multi-label, with $b={\rm c/v}= \pm 1$ denoting the conduction and valence band, respectively. 
The related eigenvectors are 
\begin{equation}\label{pses}
\psi_{\alpha v}(\mathbf{r})=\langle \mathbf{r} \vert \alpha v \rangle 
=
\frac{
e^{\imath \mathbf{k} \cdot \mathbf{r}}
}
{
\sqrt{4\pi R L}
}\, 
\left(
\begin{matrix}
1 \\
bv e^{\imath v \theta_\mathbf{k} }
\end{matrix}
\right)
\
\ ,
\end{equation}
where $\theta_\mathbf{k}$ is the polar angle of the two-dimensional wavevector $\mathbf{k}$, and $L$ denotes the nanotube length, which we assume to be the longest lengthscale in the problem, $L \rightarrow \infty$.
While for $\nu \neq 0$ the energy spectrum is gapped  (semiconducting nanotube) and near $k = 0$ is parabolic-like similarly to conventional semiconductors, for $\nu = 0$ the spectrum is gapless (metallic case), and the typical massless Dirac-cone structure is recovered. All armchair and (3n,0) zigzag SWNTs are remarkable examples of the metallic case.\cite{Ando05a}\\

{\it Phonon spectrum}. In the long-wavelength phonon limit, the transversal phonon wavevector $q_\perp$ vanishes. The SWNT phononic spectrum only depends on the wavevector $q$ along the SWNT axis, and includes {\it zone-center} and {\it zone-boundary} (ZB) modes \cite{Ando05a,Suzuura02a,Suzuura08a}. The former can be  grouped into (i) longitudinal(L) stretching modes, characterised by an  acoustic(A) branch $\omega_{q,{\rm LA}}=v_{\rm L} |q|$ with $v_{\rm L} \simeq 1.9 \times 10^4 {\rm m/s}$  and an  optical(O) branch with $\hbar \omega_{\rm LO}\simeq 0.2$ eV; (ii) breathing(Br) modes orthogonal to the nanotube surface, with a roughly $q$-independent spectrum $\hbar \omega_{{\rm Br}} \simeq 0.14 \,{\rm eV \AA }/R$; (iii) transverse(T) twisting modes, characterised by an acoustic  branch with $\omega_{q,{\rm TA}}=v_{\rm T} |q|$ with $v_{\rm T}\simeq 1.5 \times 10^4 {\rm m/s}$ and an  optical(O) branch with $\hbar \omega_{\rm TO}\simeq 0.2$ eV. In contrast ZB modes, primarily corresponding to the Kekul\'e distortions, characterized by a typical phonon energy $\hbar \omega_{\rm ZB} \simeq 0.16 \, {\rm eV}$.

{\it Electron-phonon coupling}.
As far as electron-phonon coupling is concerned, a few preliminary remarks are in order. First, while zone-center modes induce intravalley scattering, zone-boundary modes cause intervalley scattering. Secondly, not all the above modes are relevant for  our investigation. In particular, optical modes and zone-boundary modes typically become important only at very high energies, as observed, e.g., in transport measurements at high applied voltage bias~\cite{Park08a,Yao00a}. As we shall discuss in detail later, we consider here values of nonequilibrium excess energy that are much smaller than $\hbar \omega_{\rm LO},\hbar \omega_{\rm TO}$ and $\hbar \omega_{\rm ZB}$. In such a regime, only scattering with acoustic and breathing modes actually matters, whereas the contribution of optical and zone-boundary modes is definitely negligible. In fact, in Sec.\ref{sec-IV-B} we shall explicitly prove that this is true for TO and LO modes; zone-boundary modes, whose energies are comparable to O modes, are expected to have a physically negligible impact too, with the unnecessary computational drawback of coupling the two valleys. For these reasons, we shall exclude ZB modes, and  consider henceforth intravalley processes only. 
The electron dynamics thus decouples into the two valleys, and in each valley electrons scatter with each vibrational mode   $\xi={\rm LA,TA,Br,LO,TO}$.

Near the energetically relevant $\mathbf{K}$ and $\mathbf{K}^\prime$ points, the  electron-phonon coupling for each vibrational mode  is described by a $2\times 2$ matrix acting on the electronic states of the related valley. In Refs. [\onlinecite{Suzuura02a},\onlinecite{Ishikawa06a}] explicit expressions for such matrices are given in the sublattice space. Here, in order to treat the electron-phonon coupling with the Lindblad-based density-matrix formalism (see Sec.\ref{s-TA}), 
it is more suitable to switch from the sublattice basis to  the $\alpha$-basis   of the electron eigenvectors~(\ref{pses}). Then,  the electron-phonon coupling is rewritten as
\begin{equation}\label{Hprime}
\hat H^{e-ph} =  \sum_{\alpha\alpha^\prime,v,q\xi}
\left(
g^{q\xi v -}_{\alpha\alpha^\prime} \hat c^\dagger_{\alpha v} \hat b^{ }_{q\xi} \hat c^{ }_{\alpha^\prime v}
+
g^{q\xi v +}_{\alpha\alpha^\prime}  \hat c^\dagger_{\alpha v} \hat b^\dagger_{q\xi} \hat c^{ }_{\alpha^\prime v} \right) \ ,
\end{equation}
where
$
g^{q\xi v \pm}_{\alpha\alpha^\prime} = g^{q\xi v\mp *}_{\alpha^\prime\alpha}
$
describe carrier-phonon matrix entries for the carrier transition $\alpha^\prime \to \alpha$ occurring in the valley $v=\mathbf{K/K^\prime}=\pm 1$ and resulting from the absorption~($-$) or emission~($+$) of a phonon with a vibrational mode $\xi$ and wavevector $q$. Furthermore, 
$\hat c^\dagger_{\alpha v}$ ($\hat c^{ }_{\alpha v}$)  and
$\hat b^\dagger_{q\xi}$ ($\hat b^{ }_{q\xi}$) 
denote the creation (annihilation) of an electron in the $\alpha v$ single-particle states (\ref{pses}), and of a $q \, \xi$ phonon, respectively. The explicit expression for the coefficients $g^{q\xi  v  \pm}_{\alpha,\alpha^\prime}$ is given in App.\ref{App-g} for the case of a metallic SWNT. 
  
\section{Lindblad-based density-matrix formalism}\label{s-TA}

In order to investigate energy dissipation and decoherence as well as quantum-diffusion phenomena induced by the nanotube phonon bath on the otherwise phase-preserving electron dynamics, we apply  the general formalism introduced in Ref.~[\onlinecite{Rosati14e}] to the SWNT model just described.
According to such a fully quantum-mechanical treatment, the time evolution of the  single-particle density matrix 
$\rho^{v}_{\alpha_1\alpha_2} = \left\langle \hat c^\dagger_{\alpha_2 v } \hat c^{ }_{\alpha_1 v} \right\rangle$
in the $\alpha$-basis of the electronic single-particle eigenstates is given by
\begin{equation}\label{DME}
\frac{d \rho^{v}_{\alpha_1\alpha_2}}{d t} 
=
\frac{\epsilon_{\alpha_1}-\epsilon_{\alpha_2}}{\imath\hbar} \rho^{v}_{\alpha_1\alpha_2} 
+
\left.\frac{d \rho^{v}_{\alpha_1\alpha_2}}{d t}\right|_{\rm scat}\quad.
\end{equation}
In Eq.(\ref{DME}), the first term on the right-hand side describes the scattering-free propagation, with $\epsilon_\alpha$ denoting the single-particle electron eigenvalues, whereas the second term
is a non-linear scattering superoperator 
\begin{widetext}
\begin{equation}\label{DME-scat}
\left.\frac{d \rho^{v}_{\alpha_1\alpha_2}}{d t}\right|_{\rm scat} 
= 
\frac{1}{2} \sum_{\alpha^\prime\alpha^\prime_1\alpha^\prime_2,\xi} 
\left[\left(\delta_{\alpha_1\alpha^\prime} - \rho^{v}_{\alpha_1\alpha^\prime}\right)
\mathcal{P}^{\xi v}_{\alpha^\prime\alpha_2,\alpha^\prime_1\alpha^\prime_2} \rho^{v}_{\alpha^\prime_1\alpha^\prime_2} 
- 
\left(\delta_{\alpha^\prime\alpha^\prime_1} - \rho^{v}_{\alpha^\prime\alpha^\prime_1}\right) 
\mathcal{P}^{\xi v *}_{\alpha^\prime\alpha^\prime_1,\alpha_1\alpha^\prime_2} \rho^{v}_{\alpha^\prime_2\alpha_2}\right]\ +\ {\rm H.c.}\quad,
\end{equation}
\end{widetext}
expressed via generalized scattering rates $\mathcal{P}^{\xi v}_{\alpha_1\alpha_2,\alpha^\prime_1\alpha^\prime_2}$,
whose explicit form is microscopically derived from  the electron-phonon Hamiltonian~(\ref{Hprime}).
More specifically, from the general scheme described in Ref.~[\onlinecite{Rosati14e}], one obtains
\begin{equation}\label{calP}
\mathcal{P}^{\xi v}_{\alpha_1\alpha_2,\alpha^\prime_1\alpha^\prime_2} = \sum_{q \pm} 
A^{q\xi v \pm}_{\alpha_1\alpha^\prime_1}
A^{q\xi v \pm *}_{\alpha_2\alpha^\prime_2}
\end{equation}
with
\begin{equation}\label{Aalphacp}
A^{q\xi v \pm}_{\alpha\alpha^\prime} =
\sqrt{
2 \pi \left(N^\circ_{q\xi}+{1 \over 2} \pm {1 \over 2}\right)
\over 
\hbar
}
g^{q\xi v \pm}_{\alpha \alpha^\prime} D^{q\xi \pm}_{\alpha\alpha^\prime} \,  
\end{equation}
where $N^\circ_{q\xi}$ is the Bose occupation number corresponding to the phonon $q\xi$, and
\begin{equation} \label{D-def}
D^{q\xi \pm}_{\alpha\alpha^\prime} =
\lim_{\delta \to 0} 
\frac{
\exp\left\{-\left[({\epsilon_\alpha-\epsilon_{\alpha^\prime} \pm \hbar\omega_{q\xi}) / 2
\delta}\right]^2 \right\} 
}{\left(2\pi \delta^2\right)^{1/4} }    
\end{equation}
is the Gaussian regularization of the total energy conservation constraint.\footnote{As discussed in Ref.[\onlinecite{Taj09b}], the choice of this regularization function, which has no specific impact on the asymptotic system dynamics, allows for a natural time-symmetrization, crucial ingredient for the derivation of our Lindblad-like scattering superoperator.}

The fully quantum-mechanical density-matrix equation (\ref{DME}) enables us to go beyond   the conventional  Boltzmann transport equation, whose space-independent version is straightforwardly recovered  in the diagonal limit
($\rho^{v}_{\alpha_1\alpha_2} = f^{v}_{\alpha_1} \delta_{\alpha_1\alpha_2}$),\footnote{Notice that the derivation of the space-dependent Boltzmann equation goes beyond the mere diagonal limit mentioned here, and requires to perform a proper spatial coarse-graining procedure,  as discussed, e.g., in Ref.[\onlinecite{Rosati14b}].}
where the generalized scattering rates reduce to the semiclassical rates provided by the standard Fermi's golden rule 
\begin{equation}\label{P}
P^{\xi v}_{\alpha\alpha^\prime} = \mathcal{P}^{\xi v}_{\alpha\alpha,\alpha^\prime\alpha^\prime} \quad.
\end{equation}
The latter provide a qualitative information about the typical time-scale of energy dissipation versus decoherence processes induced by the various phonon modes, and will play a central role in understanding the simulated experiments presented in Sec.~\ref{s-SE}.

The average value of a generic single-particle operator $\hat a$ (with matrix entries $a^{v}_{\alpha_1\alpha_2}$ in valley $v$) can be expressed in terms of the single-particle density matrix  as
$\label{av1}
a =\sum_{v} \sum_{\alpha_1\alpha_2} \rho^{v}_{\alpha_1\alpha_2} a^{v}_{\alpha_2\alpha_1}
$.
In particular, in our investigation,   two physical quantities play a central role, namely 
the spatial carrier distribution in band $b$ ($b={\rm c/v}= \pm 1$),
\begin{equation}\label{av2}
n_b(\mathbf{r}) = \sum_{v} \sum_{\alpha_1\alpha_2} \rho^{v}_{\alpha_1\alpha_2} n^{v}_{\alpha_2\alpha_1}(\mathbf{r})  \,\delta_{b_1,b}  \delta_{b_2,b}  
\end{equation}
with  
\begin{equation}\label{av2bis}
n^{v}_{\alpha_2\alpha_1}(\mathbf{r})
=\, 
\langle \alpha_2 v \vert \mathbf{r} \rangle  
 \langle \mathbf{r} \vert \alpha_1 v \rangle \,  \quad,
\end{equation}
and the corresponding (valley-averaged) carrier momentum  distribution,
\begin{equation}\label{av3}
f_{\mathbf{k}b} = \frac{1}{2}\,\sum_{v}\sum_{\alpha_1\alpha_2} \rho^{v}_{\alpha_1\alpha_2} f^{v}_{\mathbf{k}, \alpha_2\alpha_1}   \,  \delta_{b_1,b}  \delta_{b_2,b} 
\end{equation}
with
\begin{equation}\label{av3bis}
f^{v}_{\mathbf{k}, \alpha_2\alpha_1} 
=
  \langle \alpha_2 v \vert \mathbf{k} \rangle 
  \langle \mathbf{k} \vert \alpha_1 v \rangle \,  \quad.  
\end{equation}

In particular, for the case of a metallic SWNT, which is the focus here, the above expressions reduce to
\begin{equation}\label{av4}
n_b(r_\parallel) = \frac{1}{2\pi R L}\,
\sum_v \sum_{k_1 k_2 > 0}
\rho^{v}_{k_1 b  , k_2 b  } e^{\imath (k_1 - k_2) r_\parallel}
\end{equation}
and
\begin{equation}\label{av5}
f_{kb} = \frac{1}{2}\,\sum_v \rho^{v}_{k b , k b }\ \quad,
\end{equation}
respectively.

An inspection of Eq.~(\ref{av4}) shows that a non-homogeneous spatial carrier distribution is intimately related to the presence of phase coherence between different states, $k_1 \neq k_2$. In particular, the constraint $k_1 k_2 > 0$ indicates that the only density-matrix entries contributing to the spatial distribution are those with $k_1$ and $k_2$ of equal sign.
Such feature plays a crucial role in understanding the strong suppression of carrier diffusion in metallic SWNTs, as we shall discuss in Sec.~\ref{s-SE} as well as in App.~\ref{App-IFP}.

\section{Simulated experiments}\label{s-SE}

In order to show that metallic SWNTs can be utilised as quantum-mechanical channels for the non-dispersive transmission of electronic wavepackets, we have performed a numerical solution of the Lindblad-based nonlinear density-matrix equation (LBE) in (\ref{DME}). We shall henceforth focus on the metallic case ($\nu=0$ in Eq.(\ref{epsilon})), and present results of simulated experiments, where the shape of an initially prepared  wave packet is monitored while it evolves under the effect of the phonon bath.

For any arbitrary electronic state, the density matrix  can always be written as 
\begin{equation}\label{rhoini}
\rho^{v}_{\alpha_1\alpha_2} =\rho^\circ_{\alpha_1\alpha_2} + \Delta\rho^{v}_{\alpha_1\alpha_2}\ ,
\end{equation}
where 
$\rho^\circ_{\alpha_1\alpha_2} = f^\circ_{\alpha_1} \delta_{\alpha_1\alpha_2}$
is the homogeneous equilibrium state, characterized by a Fermi-Dirac distribution 
\begin{equation}\label{FD}
f^\circ_\alpha \equiv f^\circ(\epsilon_\alpha) = \frac{1}{e^{(\epsilon_\alpha-\mu)/k_B T}+1}
\end{equation}
with chemical potential $\mu$ and temperature $T$, 
and $\Delta\rho^{v}_{\alpha_1\alpha_2}$ describes a localised excitation.
Inserting Eq.~(\ref{rhoini}) into Eq.(\ref{av4}), the spatial carrier distribution is rewritten as 
\begin{equation}\label{av6}
n_b(r_\parallel) = n^\circ_b + \Delta n_b(r_\parallel)\ ,
\end{equation}
where $n^\circ_b$ is the homogeneous equilibrium charge density and
\begin{equation}\label{Deltan}
\Delta n_b(r_\parallel) = \frac{1}{2\pi R L}\,
\sum_v \sum_{k_1 k_2 > 0}
\Delta\rho^{v}_{k_1 b  , k_2 b } e^{\imath (k_1 - k_2) r_\parallel}\quad ,
\end{equation}
is the inhomogeneous density excitation.
Similarly, the momentum carrier distribution, obtained by inserting  Eq.~(\ref{rhoini}) into (\ref{av5}), reads
\begin{equation}\label{av7}
f_{kb} = f^\circ_{kb} + \Delta f_{kb}\ ,
\end{equation}
where $f^\circ_{kb}$ is the equilibrium Fermi-Dirac distribution in (\ref{FD}) and
\begin{equation}\label{Deltaf}
\Delta f_{kb} = \frac{1}{2}\,\sum_v \Delta\rho^{v}_{k b , k b }\ .
\end{equation}
The spatial and energetic profile (e.g., Gaussian-like) of the  excitation $\Delta\rho^{v}_{\alpha_1\alpha_2}$ can in principle be generated experimentally via a properly tailored optical excitation. 
While the description of the specific optical-generation process is beyond the aim of the present paper, the localisation of the initial wave packet is a crucial aspect in our analysis.
In Ref.[\onlinecite{Rosati15a}] the excitation $\Delta \rho^{v}_{\alpha_1\alpha_2}$ was chosen to   arise from the conduction band only and, most importantly, was assumed to have purely equilibrium diagonal contributions. Here we aim to go beyond such a simplified scenario, and include nonequilibrium contributions, both in the conduction and the valence band. To this purpose, we   take an initial state described by the following intravalley density-matrix excitation:
\begin{equation}\label{Deltarho}
\Delta\rho^{v}_{\alpha_1\alpha_2} = b_1 \delta_{b_1 b_2}\, \,
C\,
e^{-\frac{1}{2}
\left(\frac{
|\epsilon_k|-E_0
}{
\Delta E
}\right)^2}
\, 
e^{-\ell |k^\prime|}
\ ,
\end{equation}
where $k = (k_1+k_2)/2$ and $k^\prime = k_1-k_2$ are the usual center-of-mass momentum coordinates, while $C$ can be regarded as a sort of excitation amplitude. Notice that the excitation (\ref{Deltarho}) is independent of the valley $v=\mathbf{K/K^\prime}=\pm 1$, and has opposite signs in the conduction ($b_1={\rm c}=+1$) and in the valence band ($b_1={\rm v}=-1$), so that no total net charge excitation is injected into the SWNT.  
The parameter $\ell$ plays the role of a delocalization length: for $\ell \to \infty$ the homogeneous case is recovered, whereas for finite values of $\ell$, an interstate phase coherence (intraband polarization) is present.
Moreover, the energetic distribution of the interband excitation is parameterized by its average energy $E_0$, often referred to as excess energy, 
 together with its standard deviation $\Delta E$. Indeed the nonequilibrium density matrix in (\ref{Deltarho}) can be regarded as the after-excitation intraband state generated by an interband laser pulse with central photon energy $\hbar\omega = 2E_0$ and pulse duration $\tau =  \hbar/2\Delta E$.

We shall focus here on the armchair (10,10) SWNT,   a metallic nanotube characterized by a breathing-mode phonon energy $\hbar\omega_{\rm Br}$ of about $20$\,meV. 
In all the simulated experiments we shall adopt as an initial condition the nonequilibrium excitation in (\ref{Deltarho}), choosing a delocalization length $\ell = 0.2\mu$m (corresponding to a FWHM value of the initial peak of about $0.4\mu$m) and an energetic broadening $\Delta E = 5$\,meV, corresponding to a laser-pulse duration $\tau \simeq 70$\,fs. We shall henceforth focus  on   the low-excitation regime, and take a value of the excitation amplitude $C$ in (\ref{Deltarho})  such as to produce a small deviation in the carrier distribution, i.e. $\Delta f_{kb} \ll 1$.

\begin{figure}[h]
	\centering
\includegraphics[width=\columnwidth]{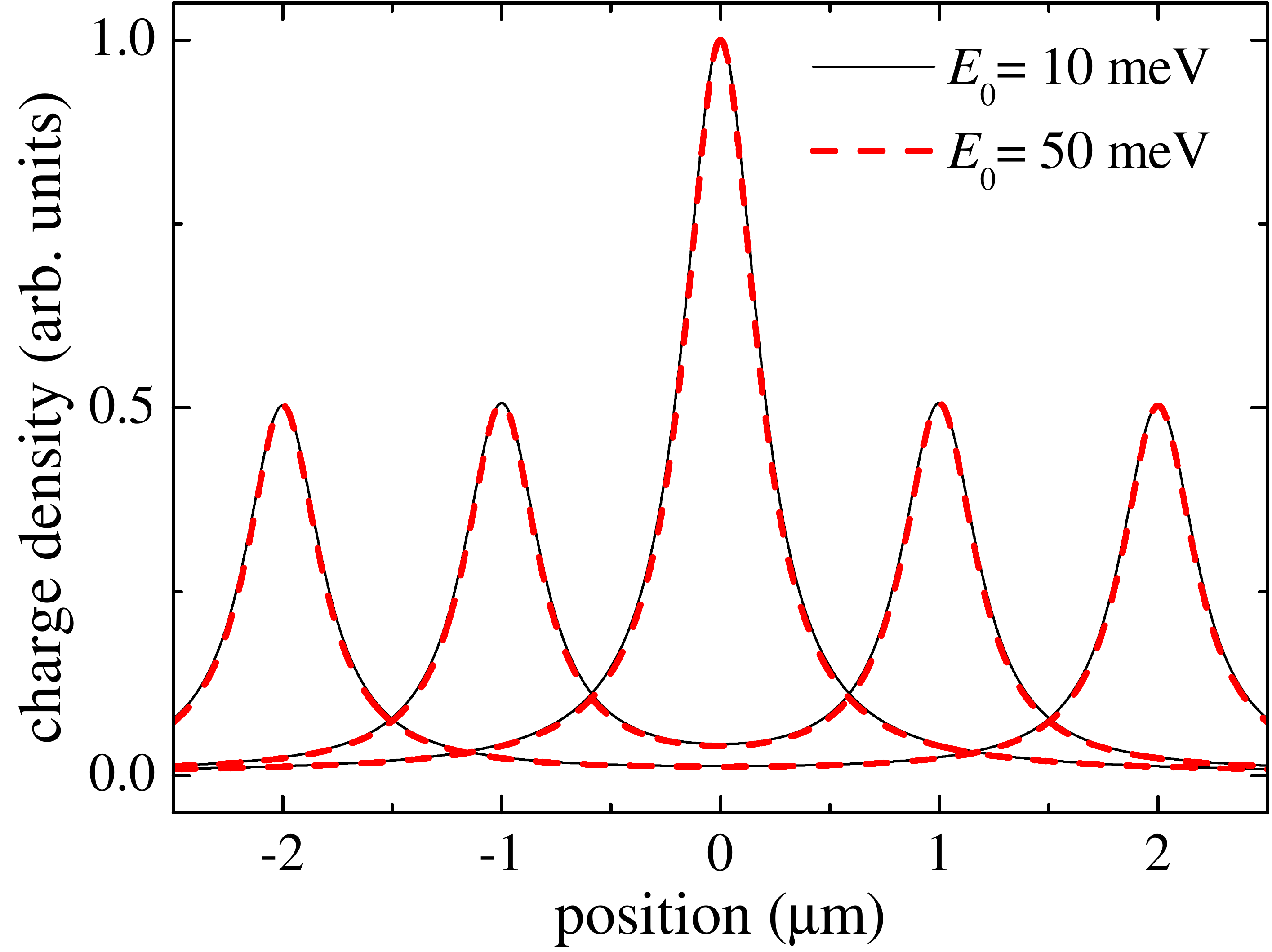}
	\caption{(Color online) 
Scattering-free dynamics of an electronic wavepacket in a metallic SWNT corresponding to the initial condition in (\ref{Deltarho}): 
conduction-band ($b = {\rm c}$) excitation charge distribution in (\ref{Deltan}) as a function of the position along the SWNT axis for an excess energy $E_0$ of $10$\,meV (solid curves) and $50$\,meV (dashed curves) at
three different times: t=0 ps (3rd peak), t=1 ps (2nd and 4th peaks), and t=2 ps (1st and 5th peaks). Note that solid and dashed lines almost coincide (see text).
}
\label{Fig1}
\end{figure}

\subsection{Scattering-free evolution}
We start our  analysis from the scattering-free propagation of the initial state in (\ref{Deltarho}) switching  off the electron-phonon coupling term in Eq.(\ref{DME-scat}). Then,  the solution of the density-matrix equation (\ref{DME}) is simply given by
\begin{equation}\label{sfs1}
\rho^{v}_{\alpha_1\alpha_2}(t) = \rho^{v}_{\alpha_1\alpha_2}(0) \, e^{-\imath (\epsilon_{\alpha_1}-\epsilon_{\alpha_2}) t/\hbar}\ ,
\end{equation}
leading to a density excitation 
\begin{equation}\label{sfs2}
\Delta n_b(r_\parallel,t) 
= 
\Delta n^{\rm r}_b(r_\parallel -   v_F t)
+
\Delta n^{\rm l}_b(r_\parallel +   v_F t)
\end{equation}
where
\begin{equation}\label{sfs3}
\Delta n^\lambda_b(r'_\parallel) 
=
\sum_{v} \sum_{k_1,k_2 \in \Omega_b^\lambda}
\Delta\rho^{v}_{k_1 b  , k_2 b }(0) 
\frac{
e^{\imath (k_1 - k_2) r'_\parallel}
}{
2\pi R L
}
\ ,
\end{equation}
with $\lambda={\rm r/l} = \pm 1$, and $b={\rm c/v}=\pm 1$. Here $\Omega_b^\lambda$ denotes
a domain defined as follows: $k_{1/2} \in \Omega_b^\lambda$ if $\lambda b k_{1/2} >0$.

 In Eq.~(\ref{sfs2}), the components $\Delta n^{\rm r/l}_b$ of the scattering-free carrier density excitations are straightforwardly  identified   as right(r)- or left(l)-moving contributions in the $b$-band, as they fulfill
\begin{equation}\label{sfs4}
\frac{d \Delta n^{\rm r/l}_b}{dt} 
=
\mp v_F \frac{d \Delta n^{\rm r/l}_b}{d r_\parallel} \quad.
\end{equation}

The splitting (\ref{sfs2}) of the carrier density evolution into right- or left-moving components is the hallmark of the well known symmetry underlying the Hamiltonian (\ref{H1}) in the case of metallic SWNTs: the  right- and left-moving electronic states (\ref{pses}) are characterised by   opposite and $k$-independent  pseudospin eigenvalues. Thus, the carrier density, which traces over the pseudospin degree of freedom (see Eqs.(\ref{av2}) and (\ref{av2bis})),  consists of oppositely propagating terms. 
Explicitly, in the $\mathbf{K}$-valley right-moving carriers have $k>0$ in the conduction band ($b=+1$) and  $k<0$ in the valence band ($b=-1$) and are all characterized by a pseudospin $+1$, whereas left-moving components have $k<0$ in the conduction band ($b=+1$) and   $k>0$ in the valence-band ($b=-1$) and are all characterized by pseudospin $-1$. The opposite pseudospin eigenvalues occur in the $\mathbf{K}^\prime$-valley.
Importantly, for a given propagation direction, all electrons are characterized by the very {\it same} velocity $v_F$, so that no wavepacket dispersion occurs. The initial charge peak thus splits into two components, which travel in opposite directions with velocity $\pm v_F$ and preserve their shape.
This is shown in Fig.~\ref{Fig1}, where the charge excitation (\ref{Deltan}) for the conduction band ($b = {\rm c}$)   is plotted as a function of the position along the SWNT axis, for an excess energy $E_0$ of $10$\,meV (solid curves) and $50$\,meV (dashed curves) at
three different times: t=0 ps (3rd peak), t=1 ps (2nd and 4th peaks), and t=2 ps (1st and 5th peaks). Similarly, an equal and opposite charge excitation arises from the valence band ($b = {\rm v}$) [not plotted here].

Importantly, as can be seen from Fig.~\ref{Fig1}, for a metallic SWNT, the shape and the propagation dynamics of the electron wave packet is nearly independent of the initial excess energy $E_0$, which is once again a peculiar feature stemming from the linearity of the band.\\

The above scenario  strongly differs from the  semiconducting SWNT case in various aspects: in the first instance, in the latter case right- and left-moving electronic eigenstates are characterized by a $k$-dependent pseudospin direction, similarly to a conventional material in the presence of spin-orbit coupling, so that the carrier density is not simply the sum of right- and left-moving terms, but also mixed terms arise. Secondly, because of the non-linearity of the band, the propagation velocity  depends on the wavevector~$k$. As a consequence, the  wave packet experiences the typical dispersion  of conventional (i.e., parabolic-band) materials, as observed in Ref.[\onlinecite{Rosati15a}]. Finally, a dependence on the initial excess energy $E_0$ arises in semiconducting SWNT.\\
 
\begin{figure}
	\centering
\includegraphics[width=\columnwidth]{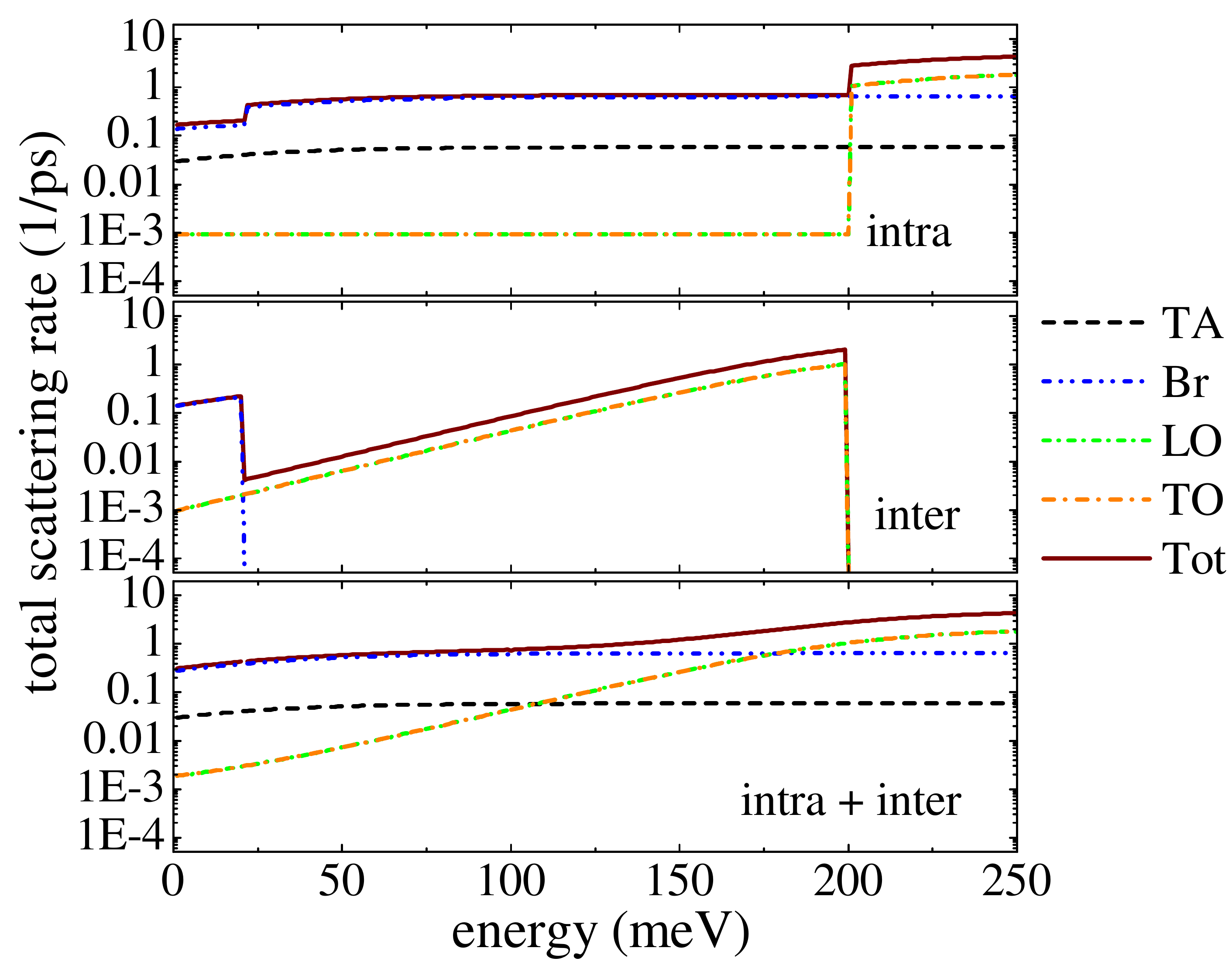}
	\caption{(Color online)
Room-temperature ($T = 300$\,K and $\mu = 0$) total scattering rates in (\ref{Gamma}) as a function of the conduction energy ($\epsilon = \hbar v_F |k|$) for intraband (top), interband (middle), and intra- plus interband scattering processes (bottom) due to the various phonon modes: $\xi$ = TA (dashed curves), Br (dotted-dotted-dashed
curves), LO (short-dotted-dashed curves), TO (dotted-dashed curves), and their sum (solid curves). Note that the $\xi = {\rm LO}$ and $\xi = {\rm TO}$ curves
almost coincide in every panel.
}
\label{Fig2}
\end{figure}

\subsection{Effects of electron-phonon coupling}
\label{sec-IV-B}
Let us now switch on the electron-phonon coupling and address the crucial question of whether and how energy dissipation and decoherence modify such an ideal dispersion-free scenario. 
To this purpose, we have performed a set of simulated experiments based on the LBE (\ref{DME}), including all the relevant phonon modes discussed in Sec.~\ref{s-PS}.

\subsubsection{Total scattering rates}
To start our analysis, a useful insight about the typical energy-relaxation time-scale is provided by the semiclassical rates $P^{\xi v}_{\alpha\alpha^\prime}$ in (\ref{P}), via the following total scattering rates
\begin{equation}\label{Gamma}
\Gamma^\xi_{k,b \to b^\prime} = \frac{1}{2}\,\sum_v \sum_{k^\prime \neq k} 
\left(1-f^\circ_{k^\prime b^\prime}\right) P^{\xi v}_{k^\prime b^\prime  , k b } 
\ ,
\end{equation}
where the generic (intravalley) transition $k b   \to    k^\prime b^\prime$ is multiplied by the Pauli-blocking factor of the final state.
The total scattering rates (\ref{Gamma}) are displayed in Fig.\ref{Fig2} as a function of the  
conduction energy ($\epsilon = \hbar v_F |k|$) for the (10,10) SWNT. 
As one can see, for both intraband and interband processes the dominant (i.e., fastest) dissipation channels are due to optical (LO and TO) and breathing (Br) phonon modes, which are expected to induce a significant energy dissipation and decoherence, in view of their strongly inelastic nature. 
In particular, for values of~$E_0$ significantly smaller than the optical-phonon energy ($\simeq 200$\,meV), the primary dissipation channel is ascribed to Br phonon modes. Furthermore, 
due to the different threshold mechanisms for intraband and interband scattering (both dictated by the phonon energy $\hbar\omega_{\rm Br} \simeq 20$\,meV), the impact of Br modes is expected to be strongly $E_0$-dependent. In any case, the total scattering rates shown in Fig.~\ref{Fig2} would suggest that the carrier-phonon scattering   induces energy dissipation and decoherence on a picosecond time-scale.
Note that LA-phonon scattering is absent for the considered (10,10) SWNT: the only available transition is $k \to k$, the so-called self-scattering.

The crucial question  to address is whether and to what extent such incoherent dynamics modifies the dispersion-free propagation scenario of Fig.~\ref{Fig1}.
Indeed, combining Eqs.~(\ref{DME}), (\ref{rhoini}), and (\ref{Deltan}), in the presence of carrier-phonon scattering the dispersion-free result in (\ref{sfs4}) is modified to
\begin{equation}\label{DME-Deltan}
\frac{d \Delta n^{\rm r/l}_b}{dt} 
=
\mp   v_F \frac{d \Delta n^{\rm r/l}_b}{d r_\parallel}
+
\left.\frac{d \Delta n^{\rm r/l}_b}{dt}\right|_{\rm scat} 
\end{equation}
with
\begin{equation}\label{DME-Deltan_scat}
\left.\frac{d \Delta n^{\lambda}_b}{dt}\right|_{\rm scat} 
=
\sum_{v} \sum_{k_1,k_2 \in \Omega_b^\lambda}
\frac{
e^{\imath (k_1 - k_2) r_\parallel}
}{
2\pi R L
}
\left.\frac{d\rho^{v}_{k_1 b  , k_2 b  }}{d t}\right|_{\rm scat}\ .
\end{equation}
and $\Omega_b^\lambda$ is defined below Eq.(\ref{sfs3}).

\begin{figure*}
	\centering
\includegraphics[width=\textwidth]{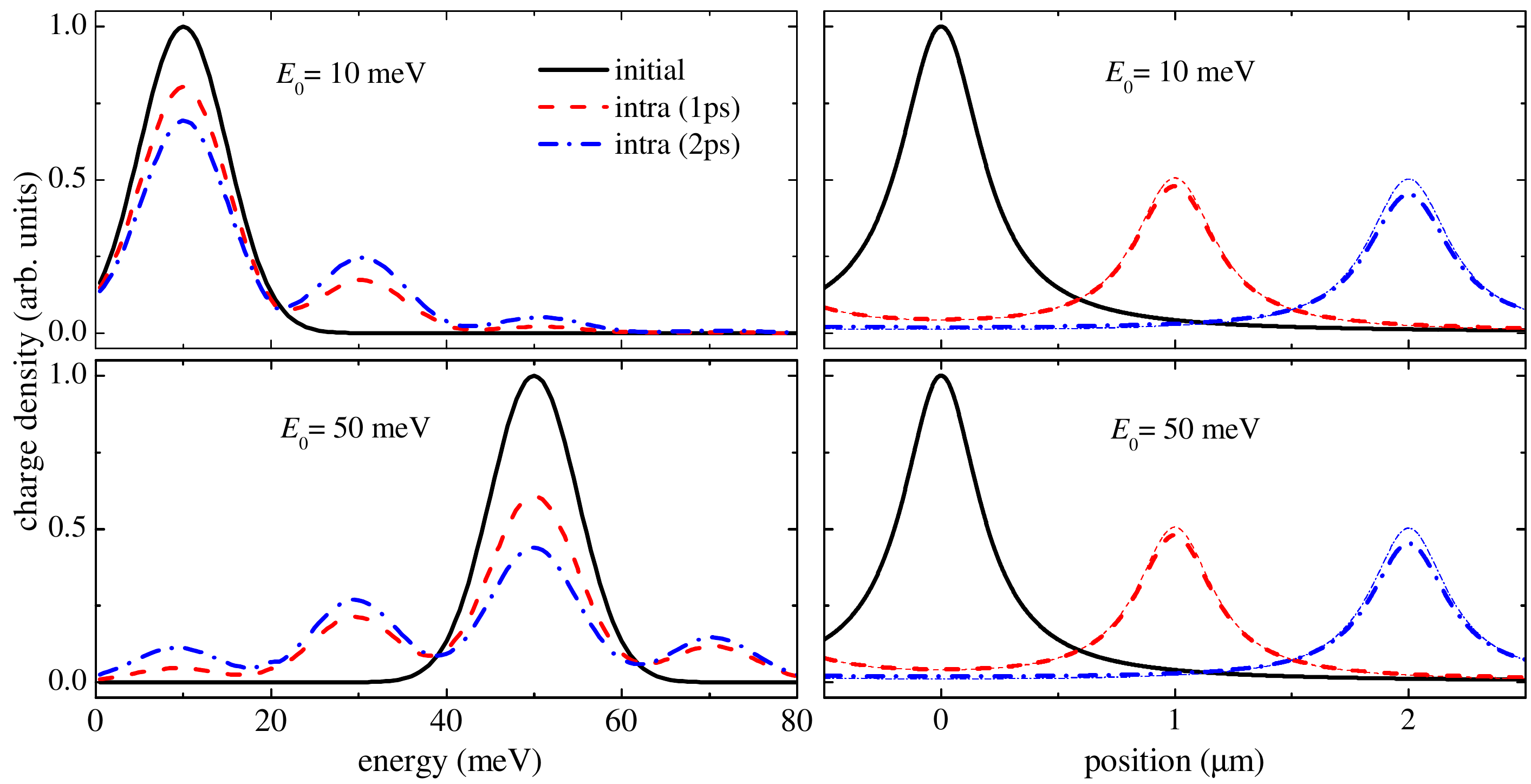}
	\caption{(Color online) 
Room-temperature ($T = 300$\,K and $\mu = 0$) dynamics of an electronic wave packet in a metallic SWNT corresponding to the initial condition in (\ref{Deltarho}) in the presence of intraband scattering only: 
conduction-band ($b = {\rm c}$) excitation charge distribution in (\ref{Deltaf}) as a function of the carrier energy $\hbar v_F |k|$ (left)
and corresponding excitation charge distribution in (\ref{Deltan}) as a function of the position along the SWNT axis (right) at three different times ($t=0 $ (solid curves), $t= 1$\,ps (dashed
curves), and $t = 2$\,ps (dash-dotted curves)) for two different excess energies: $E_0 = 10$\,meV (upper panels) and $E_0 = 50$\,meV (lower panels). 
Here all thick curves show the effects of the intraband carrier-phonon coupling accounted for by the LBE (\ref{DME}) while the thin ones in the right panels correspond to their scattering-free counterparts (see text).
} 
\label{Fig3}
\end{figure*}

\subsubsection{Intraband scattering}
Let us start by considering the case of intraband scattering processes only, where all interband dissipation channels are  switched off.
Figure \ref{Fig3} shows a direct comparison between energetic (left panels) and spatial  distributions (right panels) for  conduction band excitation carriers at different times, for two values of the excess energy, $E_0 = 10$\,meV (upper panels) and $E_0 = 50$\,meV (lower panels). The chemical potential is set here at the charge neutrality point, $\mu=0$, so that the valence band excitation distributions are equal in magnitude and opposite in sign to the conduction ones, and are not explicitly shown.
As one can see, both the energetic carrier distributions (left panels) exhibit the typical phonon-replica scenario of ultrafast energy-relaxation experiments. In particular, the nature (i.e., number of emitted Br phonons) and time-scale of the dissipation process depend on the value of the excess energy $E_0$, and agree with the intraband scattering rates reported in the upper panel of Fig.~\ref{Fig2}.
In spite of such picosecond energy-relaxation and decoherence dynamics, the spatial carrier distributions (right panels) clearly show that the electron-phonon coupling does not significantly alter the shape of the electron wavepacket with respect to the ideal  scattering-free  results (thin curves), so that (i) the propagation is essentially dispersionless up to the micrometric scale, even at room temperature, and (ii) the small diffusion effect is nearly independent of $E_0$.

\begin{figure}
	\centering
\includegraphics[width=\columnwidth]{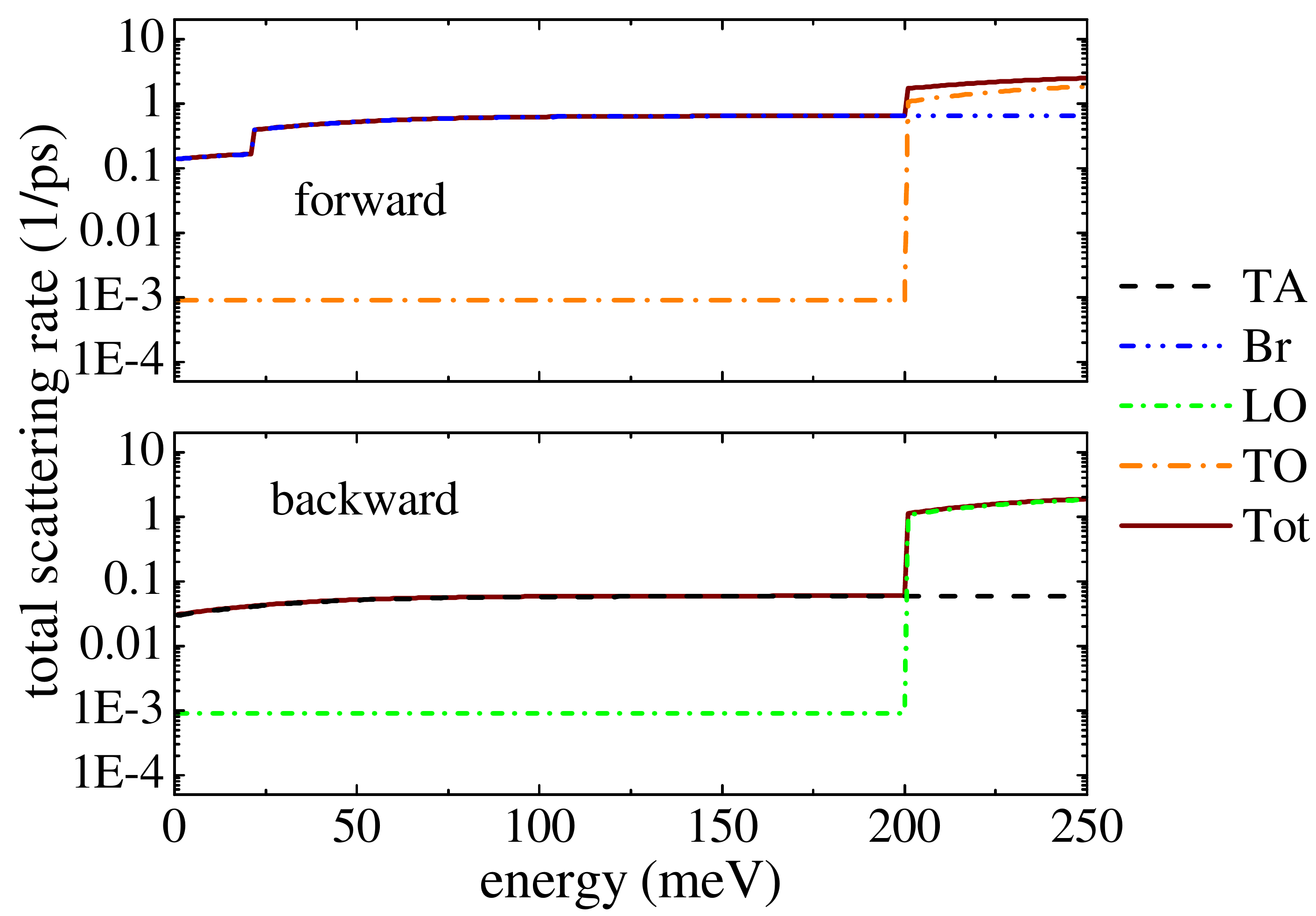}
	\caption{(Color online)
Forward-scattering (upper panel) and backward-scattering components (lower panel) of the intraband total scattering rates reported in the upper panel of Fig.~\ref{Fig2} (see text).
}
\label{Fig4}
\end{figure}

In order to understand the origin of such shape-preserving dynamics, it is worth noting that in the electron-phonon coupling a natural distinction arises  between forward and backward scattering processes, namely processes where the initial and final electronic states have the same and opposite velocity sign, respectively. 
In terms of our density-matrix formalism, since a quantum transition involves {\it two} pairs of momenta $(k_1,k_2) \to (k_1+q,k_2+q)$,   forward and backward processes can in principle interplay in Eq.~(\ref{DME-scat}). In semiconducting materials such transitions can lead to  scattering nonlocality and quantum diffusion speed-up phenomena;\cite{Rosati14b} moreover, in Luttinger liquids the forward component of the electron-phonon coupling can lead to Wentsel-Bardeen  instabilities of the electron propagator.\cite{Loss94a} 
However, in a metallic SWNT, due to the energy and momentum conservation, mixed (forward-backward) processes occupy a vanishing measure subset of the phase space, and are irrelevant. Furthermore, a detailed investigation summarized in App.~\ref{App-IFP} shows that intraband forward processes yield a negligible contribution to the scattering term in (\ref{DME-Deltan_scat}) and therefore have an extremely small impact on the wavepacket propagation. 
The wavepacket dispersion (see right panels in Fig.~\ref{Fig3}) originates mainly from backward processes. Such conclusion, obtained from a fully quantum-mechanical approach, turns out to be essentially similar to the expectation one would formulate   on a semiclassical argument based on the Boltzmann  theory.

At room temperature, the backward scattering processes may be ascribed to different phonon modes, depending on the type of SWNT: for armchair SWNT, like the (10,10) one, they are due to TA modes only, whereas for   zigzag SWNTs they are due to Br as well as to LA modes. 
We stress that also LO modes induce backward processes; however, due to their high phonon energy \cite{Yao00a}, in the simulated experiments of Fig.~\ref{Fig3} their impact is extremely negligible. In turn, this also confirms that the neglect of the zone-boundary modes --whose energy is comparable to the optical modes-- is a good approximation. 

The scenario described so far is confirmed by the forward-versus-backward total scattering rates reported in Fig.~\ref{Fig4}.
As anticipated, the intraband scattering rates in the upper panel of Fig.~\ref{Fig2} are dominated by forward processes (see upper panel in Fig.~\ref{Fig4}) which, in turn, are dominated by Br phonon modes.
In contrast, the total scattering rate due to backward processes (see lower panel in Fig.~\ref{Fig4}) is due to TA modes only, and is at least one order of magnitude smaller compared to the forward one.
Recalling that the diffusion of an electronic wavepacket in a metallic SWNT is mainly determined by backward processes (see App.~\ref{App-IFP}) and that the latter are characterized by a much longer time-scale, we are then able to explain the apparent discrepancy in Fig.~\ref{Fig3} between the energy-relaxation (left panels) and the quantum diffusion time-scale (right panels).
Moreover, the fact that quantum diffusion is mainly determined by backward processes, and that the latter involve TA phonons only, explains well how the diffusion dynamics (right panels) is basically independent of the excess energy $E_0$.

\subsubsection{Effects of interband scattering}
\begin{figure*}
	\centering
\includegraphics[width=\textwidth]{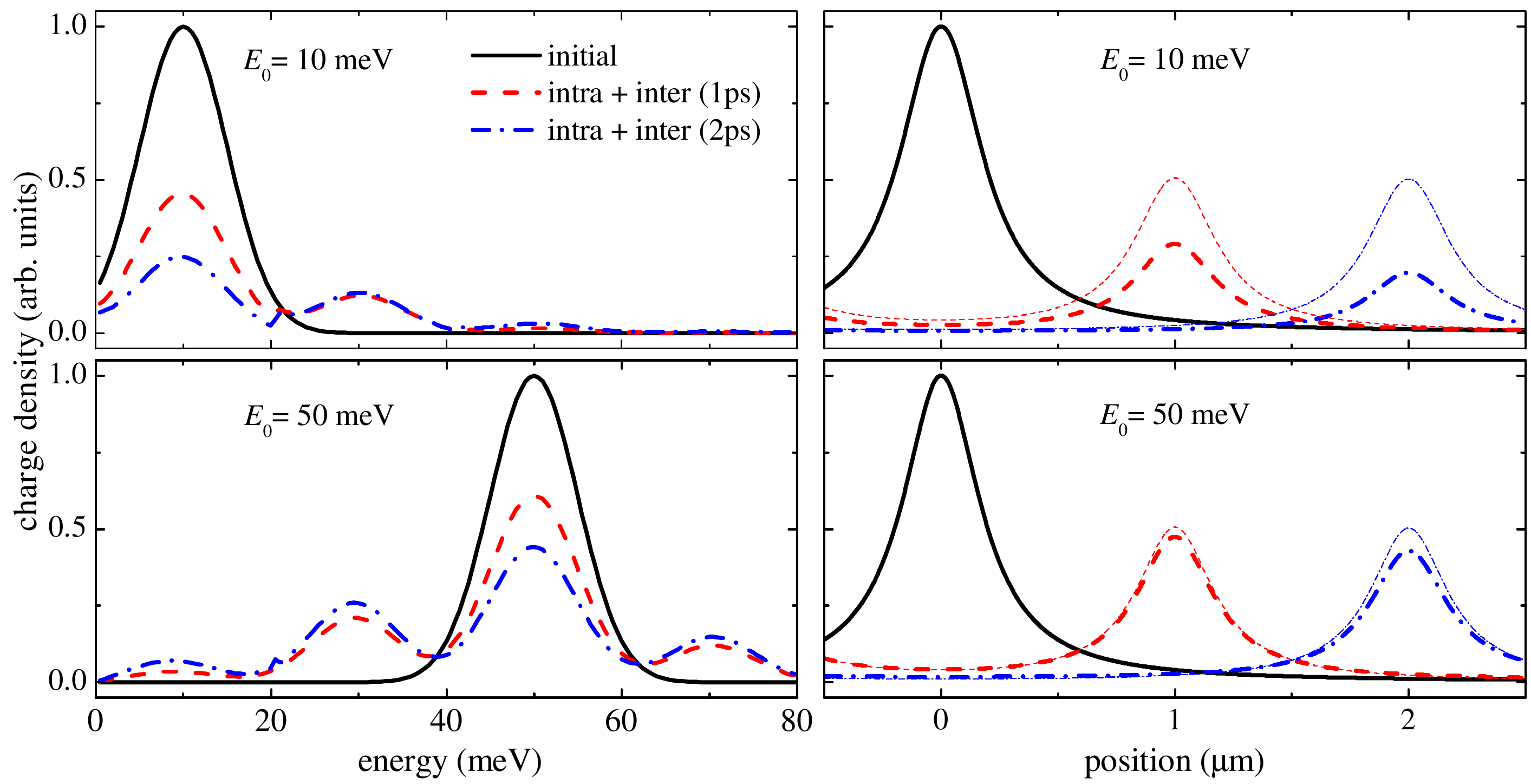}
	\caption{(Color online) 
Same quantities as in Fig.~\ref{Fig3}, but in the presence of both intra- and   interband carrier-phonon scattering. 
While for small excess energy ($E_0=10{\rm meV}$) interband scattering processes affect  both the energy and space distributions (compare upper panels with those of Fig.\ref{Fig3}), for higher excess energy ($E_0=50{\rm meV}$) interband scattering has a negligible impact (compare lower panels with those of Fig.\ref{Fig3}).
} 
\label{Fig5}
\end{figure*}

As a second step, we have included also interband carrier-phonon scattering, and analyzed how the  simulated experiments of Fig.~\ref{Fig3} are modified by the presence of such processes.
Figure \ref{Fig5} shows again a direct comparison between energy-relaxation (left panels) and spatial-diffusion dynamics (right panels), for the same two values of $E_0$. 
While for $E_0 =10$\,meV the presence of interband scattering induces strong modifications with respect to the intraband results of Fig.~\ref{Fig3}, for $E_0 = 50$\,meV, the effect of interband coupling is hardly visible, both in terms of the energetic and the spatial carrier distributions.
Indeed, an inspection of the interband total scattering rates reported in the central panel of Fig.~\ref{Fig2} shows that for the considered values of $E_0$ the most efficient (i.e., fastest) interband scattering channel is again ascribed to Br phonon modes; however, such a picosecond scattering mechanism is active only for carrier energies smaller than $\hbar\omega_{\rm Br}$.
Moreover,  in addition to an energetic carrier redistribution, the presence of interband transitions leads to a progressive decay of the initial excitation charge $\Delta n_b(r_\parallel)$ in (\ref{Deltan}) via an interband charge transfer, which can be regarded as a net phonon-mediated electron-hole recombination process.
The resulting loss of conduction electrons may affect the nearly dispersion-free scenario of Fig.~\ref{Fig3}. However, its impact is directly related to the effective time-scale of such phonon-induced interband transfer, which, in turn, depends on the fraction of below-threshold ($\epsilon < \hbar\omega_{\rm Br}$) electrons, and therefore on the value of $E_0$.

\begin{figure}
	\centering
\includegraphics[width=\columnwidth]{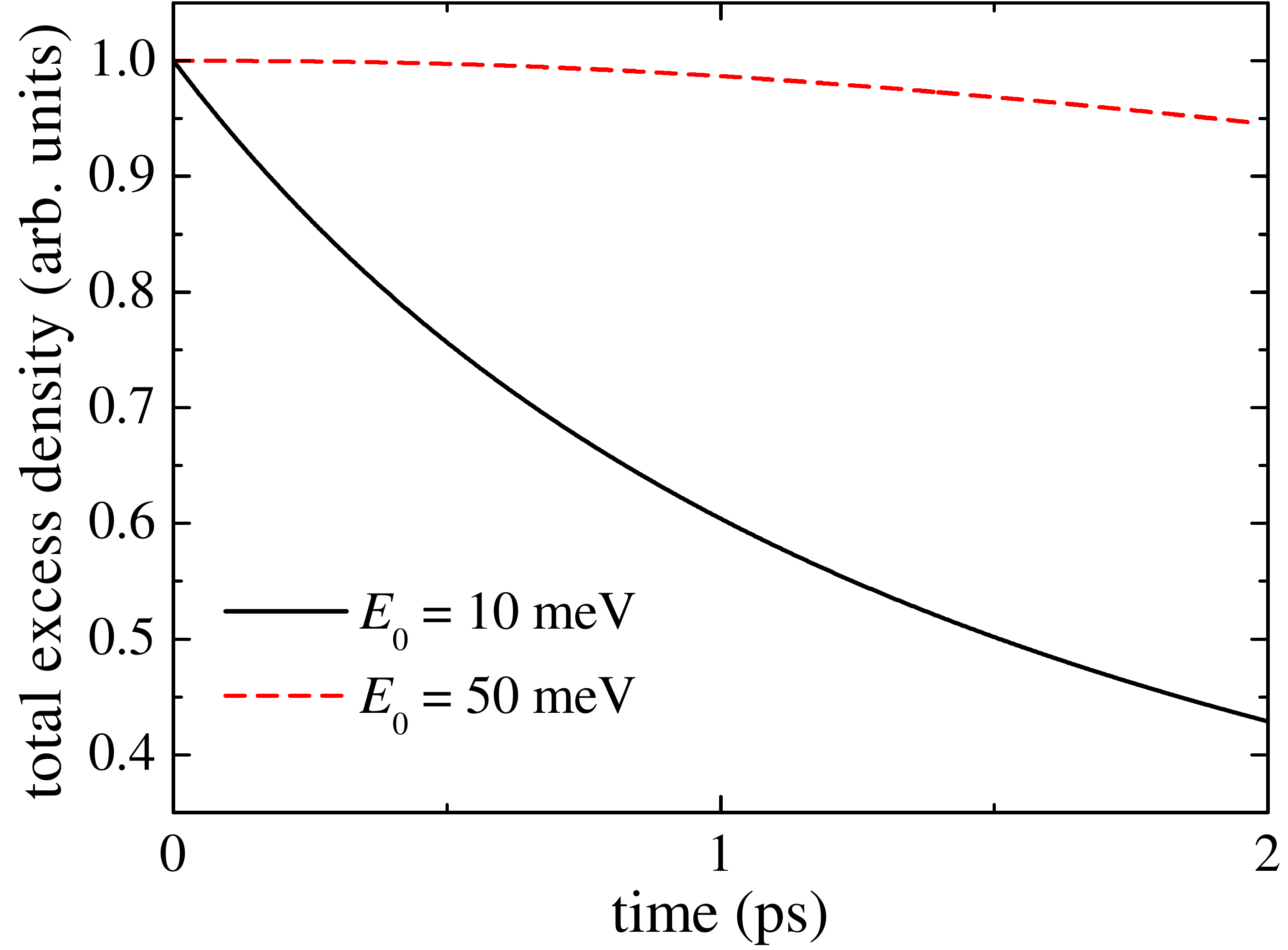}
	\caption{(Color online) 
Time evolution of the total excess density corresponding to the two simulated experiments of Fig.~\ref{Fig5}: $E_0 = 10$\,meV (solid curve) and $E_0 = 50$\,meV (dashed curve) (see text).
} 
\label{Fig6}
\end{figure}

Such highly nontrivial interplay between the conduction-band energy redistribution and electronic loss due to phonon-induced interband transfer is fully confirmed by the two simulated experiments of Fig.~\ref{Fig5}.
For a given initial excitation peak with an excess energy $E_0$ (see solid curves in the left panels), the conduction electrons experience a sequence of Br-phonon emissions and/or absorptions, giving rise to corresponding phonon replica in the excitation charge distribution. The resulting time-scale of interband scattering is then related to the number of emitted phonons needed to enter the below-threshold energy region, and thus increases for increasing values of $E_0$.
Such a behavior is fully confirmed by the time evolution of the total excess density reported in Fig.~\ref{Fig6}, which shows that, by increasing   $E_0$ from $10$ to $50$\,meV, the net interband carrier transfer is reduced by more than one order of magnitude.
The relevant conclusion is that for excess energies $E_0 > \hbar\omega_{{\rm Br}}$ the room-temperature wavepacket propagation is again essentially dispersionless up to the micrometric scale also in the presence of interband scattering.

To conclude this subsection, we observe that the energy and space carrier distributions shown in  Figs.\ref{Fig3} and \ref{Fig5} have been chosen as the most suitable quantities to specifically address the problem of the wavepacket dispersion. The density matrix obtained by solving Eq.(\ref{DME-scat}) --or equivalently its related Wigner function-- encodes further information, however, its description is beyond the purposes of the present paper.   A similar analysis, carried out on parabolic quantum wires within the Wigner function formalism can be found e.g. in Ref.~[\onlinecite{Herbst03a}].

\subsubsection{Effects of the chemical potential}
\begin{figure}
	\centering
\includegraphics[width=\columnwidth]{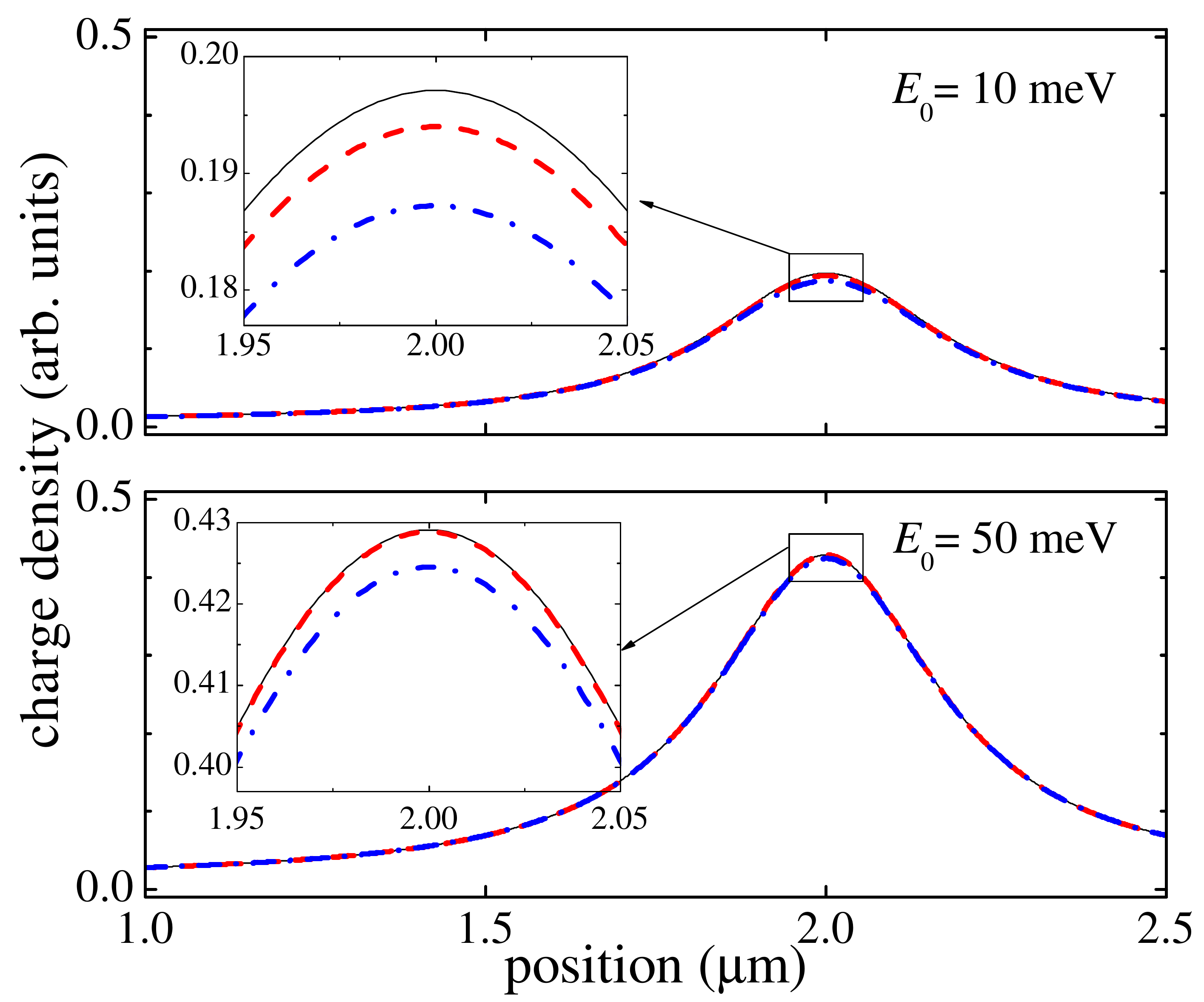}
	\caption{(Color online)
Snapshot of the wavepacket spatial distribution after $2$\,ps at room temperature for different values of the chemical potential: $\mu=0$ (thin solid), $\mu=40 {\rm meV}$ (dashed), and $\mu= -80 {\rm meV}$ (dashed-dotted). Upper and lower panels refer to $E_0 = 10$\,meV   and $E_0 = 50$\,meV, respectively. A very weak dependence is observed on the chemical potential, which becomes appreciable only when zooming near the peaks, as shown by the two insets (see text).
} 
\label{Fig7}
\end{figure}

So far, all the described simulated experiments (see Figs.~\ref{Fig3} and \ref{Fig5}) have been performed for a value of the chemical potential $\mu$ corresponding to the charge neutrality point: $\mu = 0$. 
We now want to discuss the effects of the chemical potential on the wavepacket propagation. In particular, one would expect that, as the chemical potential is increased or decreased, the change in the occupation of initial and final electronic states available for electron-phonon scattering alters the relative weight of intra- and interband scattering contributions (see Eq.~(\ref{Gamma})). Furthermore, one expects that, away from the charge neutrality point $\mu=0$, the magnitudes of conduction- and valence-band carriers become different. 

To analyze these effects, the simulated experiments of Fig.~\ref{Fig5} have been repeated with varying the value of the chemical potential. Figure~\ref{Fig7} shows snapshots of the wave-packet spatial distribution taken $2$\,ps after the initial condition for different values of $\mu$. The upper and lower panels refer to the same two excess energy values of Figs.~\ref{Fig1}, \ref{Fig3}, and \ref{Fig5}, namely $E_0 = 10$\,meV and $E_0 = 50$\,meV,  respectively.
As one can see, the wavepacket spatial profile exhibits a weak dependence on~$\mu$. Similarly,  a small modification was found on the  energy-relaxation process, and has not been reported here.
Surprisingly, such independence occurs even for small values of excess energy $E_0$, where the interband contribution has been shown to modify the intraband results, as discussed above (see Figs.~\ref{Fig5} and \ref{Fig6}). 
Furthermore, only a minor difference turns out to arise for $\mu \neq 0$ between the magnitudes of the conduction and valence band carrier distributions: the relative difference of the maximum heights is less than $2 \%$.

In order to explain such seemingly counter-intuitive behavior, it is useful to describe the effect of the relevant interband scattering channel, namely Br phonon modes, via a simple two-level toy model, which involves just one single conduction (${\rm c}$)  and a single valence (${\rm v}$) state. 
More specifically, we shall denote with 
$\epsilon_{\rm c/v} = \pm  \hbar\omega_{{\rm Br}}/2$ the corresponding energy levels, with $f_{\rm c/v}$ the corresponding electron population, and with $P_{\rm cv} = W N^\circ_{{\rm Br}}$ and $P_{\rm vc} = W \left(N^\circ_{{\rm Br}}+1\right)$ the interlevel absorption and emission rates, respectively ($N^\circ_{{\rm Br}}$ denoting the Breathing mode Bose occupation number).
Within the conventional semiclassical picture, the time evolution of the electron populations is described by the following Boltzmann-like equation: 
\begin{equation}\label{BTE}
\frac{d f_{\rm c}}{d t} 
=
(1-f_{\rm c}) P_{\rm cv} f_{\rm v} - (1-f_{\rm v}) P_{\rm vc} f_{\rm c}
=
-\frac{d f_{\rm v}}{d t}\ .
\end{equation}
Writing the two electron populations as 
$f_{\rm c/v} = f^\circ_{\rm c/v} \pm \Delta f$ 
($\Delta f$ denoting the deviation from the thermal-equilibrium distribution $f^\circ_{\rm c/v}$) and neglecting quadratic terms in $\Delta f$, Eq.~(\ref{BTE}) 
reduces to
\begin{equation}\label{LBTE}
\frac{d \Delta f}{d t} = - \Gamma \Delta f \quad,
\end{equation}
where
\begin{equation}\label{Gammamu}
\Gamma(\mu) = \left(P_{\rm cv}+P_{\rm vc}\right)
\left(1 - 
\frac{
\Delta f^\circ_{\rm vc}(\mu)
}{
2 N^\circ_{{\rm Br}} +1
}
\right)\ ,
\end{equation}
with $\Delta f^\circ_{\rm vc}(\mu) = f^\circ_{\rm v}(\mu)-f^\circ_{\rm c}(\mu)$ denoting the difference between valence and conduction Fermi-Dirac functions.
Equation (\ref{LBTE}) shows that the initial excess population $\Delta f$ undergoes an exponential-decay dynamics according to the $\mu$-dependent decay rate in (\ref{Gammamu}), whose relative change with respect to the $\mu = 0$ case is given by
\begin{equation}\label{DeltaGammamu}
\Delta\Gamma(\mu) \equiv \frac{\Gamma(\mu)-\Gamma(0)}{\Gamma(0)} \quad,
\end{equation}
i.e., a positive, finite, and symmetric function of $\mu$. This implies that the decay rate $\Gamma$ in (\ref{Gammamu}) is minimal for $\mu = 0$ and increases with $|\mu|$, reaching a saturation value for $|\mu| \to \infty$.
It is, however, straightforward to verify that for the parameters of the simulated experiments reported in Fig.~\ref{Fig7}, namely $T =300$\,K, $|\mu| = 40$\,meV, and $\hbar\omega_{{\rm Br}} \simeq 20$\,meV, the relative change in (\ref{DeltaGammamu}) is only $3$\%.
Moreover, also for $|\mu| \to \infty$ the latter never exceeds its limiting value of about $7$\%.
We emphasize that such extremely weak $\mu$ dependence is ascribed to the room-temperature regime considered here. Indeed, at $T = 77$\,K the $3$\% value obtained at room temperature increases to about 140\%, which implies a strong $\mu$ dependence in the low-temperature limit, as expected.
Regardless of the specific $\mu$ dependence, our analysis shows that the impact of interband carrier-phonon scattering is always minimum for $\mu = 0$.

\section{Summary and conclusions}\label{s-SC}

We have investigated in detail  the impact of carrier-phonon coupling on the dynamics of an electron wavepacket propagating in metallic SWNTs, utilizing a recently developed density-matrix approach\cite{Rosati14e} that enables us to account for both energy dissipation and  decoherence effects. The recent study in Ref. \onlinecite{Rosati15a} has been extended in various aspects in this paper: (i) we have considered  the case of nonequilibrium carrier distributions; (ii) we have included interband carrier-phonon coupling; (iii) we have analyzed the effects of dissipation and related decoherence phenomena on the wavepacket energetic distribution; and (iv) we have discussed the effects of the chemical potential.  Based on our analysis, we can extend the conclusion that in  metallic SWNTs the shape of the wavepacket is essentially unaltered, even in the presence of intraband as well as interband electron-phonon coupling, up to micrometer distances at room temperature. More specifically, our investigation has shown that, in spite of a significant energy-dissipation and decoherence dynamics, electronic diffusion in metallic systems is extremely small as well as nearly independent of the wavepacket energetic distribution, namely excess energy and chemical potential. 
Our results thus indicate that metallic SWNTs constitute a promising platform to realize quantum channels for the non-dispersive transmission of electronic wave packets.

\begin{acknowledgments}

We are grateful to Massimo Rontani for stimulating and fruitful discussions.
We gratefully acknowledge funding by the Graphene@PoliTo laboratory of the Politecnico di Torino, operating within the European FET-ICT Graphene Flagship project (www.graphene-flagship.eu). F.D. also  acknowledges financial support from Italian FIRB 2012 project HybridNanoDev (Grant No.RBFR1236VV).

\end{acknowledgments}

\appendix

\section{Electron-phonon coupling coefficients}
\label{App-g}
In this appendix we provide the explicit expression for the $g^{q \xi v \pm}_{\alpha \alpha^\prime}$ coefficients appearing in the electron-phonon coupling Hamiltonian (\ref{Hprime}), focusing  on the case of a metallic SWNT [$\nu=0$ in Eq.(\ref{epsilon})]. The coefficients can be obtained  from the $2\times 2$   electron-phonon matrices given for   the sublattice basis in Refs.[\onlinecite{Suzuura02a},\onlinecite{Ishikawa06a}],    by changing   to the eigenvector basis~$\alpha$ defined in Eq.(\ref{pses}). Recalling that in $g^{q \xi v \pm}_{\alpha \alpha^\prime}$  multi-labels for   the electronic states are $\alpha=(k,b)$ and $\alpha^\prime=(k^\prime,b^\prime)$, the conservation of total momentum  implies that the $g^{q\xi v \pm}_{\alpha\alpha^\prime}$ exhibit the form
\begin{equation}
g^{q\xi v \pm}_{\alpha\alpha^\prime}= g^{\xi  v}_{k,k\pm q;b,b^\prime} \, \delta_{k\pm q, k^\prime} \,  \quad, \label{g-mom-cons}
\end{equation}
where the $g^{\xi   v}_{k,k\pm q;b, b^\prime}$ acquire the following expressions
\begin{widetext} 

\begin{eqnarray}
g_{k,k\pm q;b,b^\prime}^{{\rm LA} v} &=& - \sqrt{\frac{\hbar \, |q|}{2 N M v_{\rm L} }} \, e^{-i v (\theta-\theta^\prime)/2} 
\left[ g_1 f^{\rm s}(|q|) \left( \frac{1+b b^\prime}{2} \cos\frac{\theta-\theta^\prime}{2} +i\, v \, \frac{1- b b^\prime}{2}  \, \sin\frac{\theta-\theta^\prime}{2}\right) \right. + \hspace{1cm}\nonumber \\
& & \hspace{5cm} \left. +g_2 \, v\, \frac{b+b^\prime}{2} \, \cos\left(3\eta+ \frac{\theta+\theta^\prime}{2}\right)  -g_2 \, i \, \frac{b-b^\prime}{2}   \sin \left( 3\eta+ \frac{\theta+\theta^\prime}{2}\right) \right] \label{g-LA} \\
& &    \nonumber 
\end{eqnarray}
\begin{eqnarray}
g_{k,k\pm q;b,b^\prime}^{{\rm TA} v} &=&   \sqrt{\frac{\hbar \, |q|}{2 N M v_{\rm T} }} \, e^{-i v (\theta-\theta^\prime)/2} 
g_2 \, \left[   \, v\, \frac{b+b^\prime}{2}     \sin\left( 3\eta+\frac{\theta+\theta^\prime}{2}\right) +    \, i \, \frac{b-b^\prime}{2}      \cos\left(3\eta +\frac{\theta+\theta^\prime}{2}\right)   \right] \,   \label{g-TA}\hspace{1.6cm}
\end{eqnarray}
\begin{eqnarray}
g_{k,k\pm q;b,b^\prime}^{{\rm LO} v} &=& - \frac{2^{3/2}\,  \hbar v_F}{a_0^2}  \sqrt{\frac{\hbar }{2 N M \omega_{\rm LO} }} \, e^{-i v (\theta-\theta^\prime)/2}   \mbox{sgn}(q) \left[ \frac{b+ b^\prime}{2}  \, i \, v  \,  \cos\left(\frac{\theta+\theta^\prime}{2}\right)    \,  + \, \, \, \frac{b- b^\prime}{2}  \,    \,  \sin\left( \frac{\theta+\theta^\prime}{2}\right)  \right] \,     \label{g-LO}
\end{eqnarray}
\begin{eqnarray}
g_{k,k\pm q;b,b^\prime}^{{\rm TO} v} &=& -  \frac{2^{3/2}\,  \hbar v_F}{a_0^2} \sqrt{\frac{\hbar }{2 N M \omega_{\rm TO} }} \, e^{-i v (\theta-\theta^\prime)/2}    \mbox{sgn}(q) \left[ \frac{b+ b^\prime}{2}  \,  i \, v \,    \, \sin\left(\frac{\theta+\theta^\prime}{2}\right)   \,  - \, \, \, \frac{b- b^\prime}{2}  \,  \,   \cos\left(\frac{\theta+\theta^\prime}{2}\right)   \right] \,     \label{g-TO}
\end{eqnarray}
\begin{eqnarray}
g_{k,k\pm q;b,b^\prime}^{{\rm Br} v} &=&    \frac{1}{R} \sqrt{\frac{\hbar }{2 N M \omega_{\rm Br} }} \, e^{-i v (\theta-\theta^\prime)/2}  \left[ g_1 f^{\rm s}(|q|)  \left( \frac{1+b b^\prime}{2} \cos\frac{\theta-\theta^\prime}{2} +i\, v \, \frac{1- b b^\prime}{2}  \, \sin\frac{\theta-\theta^\prime}{2}\right) \right. + \nonumber \\
& & \hspace{4.5cm} \left. -g_2 \, v\, \frac{b+b^\prime}{2}  \cdot\!\!\!  \cos\left(3\eta +\frac{\theta+\theta^\prime}{2}\right) +   g_2 \, i \, \frac{b-b^\prime}{2}    \sin\left(3\eta+ \frac{\theta+\theta^\prime}{2}\right)  \right] \quad,   \label{g-Br}
\end{eqnarray}

\end{widetext}
where $\theta=\mbox{sgn}(k)\, \pi/2 $ and $\theta^\prime= \mbox{sgn}(k\pm q)\, \pi/2$ are shorthand notations for the polar angles $\theta_{k}$ and $\theta_{k\pm q}$ of the electron wavevectors $k$ and $k\pm q$ in the metallic SWNT. In the above equations, $N$ denotes the number of unit cells, $R$  the nanotube radius, $M=19.9 \times 10^{-27} {\rm kg}$   the mass of a carbon atom~\cite{Ishii07a},  $a_0=1.44 {\rm \AA}$ is the lattice spacing, and $\eta$ is the SWNT chirality angle (e.g. $\eta=0$ for zigzag and $\eta=\pi/6$ for armchair SWNT~\cite{Ando05a}). The values for $v_{\rm L}, v_{\rm T}, \omega_{\rm TO}, \omega_{\rm LO}$ and $\omega_{\rm Br}$ are given in Sec.\ref{s-PS}.  
Furthermore, $g_1 = 30$\,eV and $g_2 = 1.5$\,eV are the coupling constants related to deformation potential and bond-length change,\cite{Suzuura02a} respectively, while $f^{\rm s}(|q|)$ denotes the screening function given in Ref.~\onlinecite{vonOppen09a}. .  

The hermiticity of the electron-phonon coupling (\ref{Hprime}) ensures $g^{\xi v}_{k,k\pm q;b,b^\prime}= g^{\xi v *}_{k \pm q,k;b^\prime,b}$, whereas   the additional relation
$
 g^{\xi v}_{k,k\pm q; b,b^\prime}= g^{ \xi -v *}_{-k,-k\mp q; b,b^\prime}$ stems from  time-reversal symmetry.\\

\section{Analysis of intraband forward scattering processes}\label{App-IFP}

In this Appendix we show that in a metallic SWNT the electron diffusion dynamics is not affected by intraband forward carrier-phonon scattering. To begin with, a comment is in order here. 
At the level of the electron-phonon Hamiltonian, each term in Eq.(\ref{Hprime}) can be written as a sum of forward and backward processes, where a forward (backward) contribution can be defined as a quantum-mechanical transition where the electron group velocity in the final state $\alpha$ has the same (opposite) direction as the one in the initial state $\alpha^\prime$. However, in our approach based on the density matrix,   {\it two} states are involved in $\rho^{v}_{\alpha_1,\alpha_2}$, and quantum-mechanical transitions $(\alpha_1,\alpha_2) \to (\alpha_1^\prime,\alpha_2^\prime)$ may in general mix backward and forward Hamiltonian contributions. For these reasons, we shall utilize here  the term 'forward' ('backward') for those processes where the group velocity is preserved (changed) for {\it both} density-matrix indices.  
A similar distinction can be made between intra- and interband processes, and we shall refer to intraband transitions as the ones where both initial and final states are in the same band. In terms of the above definitions, the case of intraband forward transitions characterizes processes where the sign of both carrier wavevectors $k_1$ and $k_2$ is preserved. We shall now argue that they do not contribute to the spatial electronic diffusion.

To this purpose, we first consider the structure of the non-linear scattering superoperator~(\ref{DME-scat}) and focus on the low-excitation regime considered in our simulated experiments. We note that, by inserting Eq.~(\ref{rhoini}) into Eq.~(\ref{DME-scat}) and neglecting quadratic terms in $\Delta\rho^{v}_{\alpha_1\alpha_2}$, the original scattering term reduces to the following linear superoperator
\begin{widetext}
\begin{equation}\label{DME-scat-lin}
\left.\frac{d \rho^{v}_{\alpha_1\alpha_2}}{d t}\right|_{\rm scat} 
= 
\frac{1}{2} \sum_{\alpha^\prime_1\alpha^\prime_2,\xi} 
\left(
\overline{\mathcal{P}}^{\xi v}_{\alpha_1\alpha_2,\alpha^\prime_1\alpha^\prime_2} 
\Delta\rho^{v}_{\alpha^\prime_1\alpha^\prime_2} 
- 
\overline{\mathcal{P}}^{\xi v *}_{\alpha^\prime_1\alpha^\prime_1,\alpha_1\alpha^\prime_2} 
\Delta\rho^{v}_{\alpha^\prime_2\alpha_2}\right)\ +\ {\rm H.c.} \quad,
\end{equation}
\end{widetext}
with effective (i.e., $\mu$-dependent) scattering rates
\begin{equation}\label{calPbar}
\overline{\mathcal{P}}^{\xi v}_{\alpha_1\alpha_2,\alpha^\prime_1\alpha^\prime_2} 
=
\left(1-f^\circ_{\alpha_1}\right)
\mathcal{P}^{\xi v}_{\alpha_1\alpha_2,\alpha^\prime_1\alpha^\prime_2} 
+
f^\circ_{\alpha_1}
\mathcal{P}^{\xi v *}_{\alpha^\prime_1\alpha^\prime_2,\alpha_1\alpha_2}\ .
\end{equation}
In the case  of intraband forward scattering processes, the  effective rates $\overline{\mathcal{P}}$ in Eq.(\ref{calPbar}) can take a simpler expression.
Indeed for intraband processes, the coefficients $g^{\xi v}_{k,k\pm q; b,b^\prime}$ appearing in Eq.(\ref{g-mom-cons}) further simplify to
a form $
g^{\xi v}_{k,k\pm q; b,b^\prime} = g^{\xi v}_{k,k\pm q;b} \, \delta_{b b^\prime}\label{g-cond-right}
$.
Moreover, when only intraband forward scattering is considered, the above $g^{\xi v}_{k,k\pm q;b}$ turn out to acquire an expression that is independent of the magnitude of $k$, and that we shall denote as $g^{q\xi v}_b$. 
This can easily be seen by focusing on an illuminating example. 

Let us consider, for instance,  the   conduction band ($b={\rm c}$) and right-moving electrons ($\lambda={\rm r}$), as illustrated by the figures shown in Sec.\ref{s-SE}. 
In this case, a direct evaluation of Eqs.(\ref{g-LA}) to (\ref{g-Br}) for $b=b^\prime = {\rm c}$ and for forward scattering 
(i.e., $k,k\pm q \in \Omega_{{\rm c}}^{{\rm r}}$ corresponding to $\theta=\theta^\prime=\pi/2$), and the use of Eq.(\ref{g-mom-cons}) reveal  that the $g^{q\xi v \pm}_{\alpha\alpha^\prime} $ take  the simple form 
\begin{equation}
g^{q \xi v \pm}_{\alpha\alpha^\prime} = g^{q\xi v}_{{\rm c}} \, \delta_{k\pm q, k^\prime} \, \, \delta_{b b^\prime} \, \delta_{b {\rm c}}\quad. \label{g-cond-right-bis}
\end{equation}
Furthermore, as a result of the linearity of the band in the metallic SWNT, in this case the energy-conservation function in (\ref{D-def})  also  becomes independent of the magnitude of $k$:
\begin{equation}\label{D-cond-right}
D^{q \xi \pm}_{\alpha\alpha^\prime} = 
D^{q\xi}_{{\rm c}} \doteq
\lim_{\delta \to 0} 
\frac{
\exp\left\{-\left[    \hbar (v_F q -  \omega_{q\xi}) / 2
\delta \right]^2 \right\} 
}{\left(2\pi \delta^2\right)^{1/4} }    \quad.
\end{equation} 
Inserting Eqs.(\ref{g-cond-right-bis}) and (\ref{D-cond-right}) into Eq.(\ref{Aalphacp}), one obtains
\begin{equation}
\label{Aalphacp-cond-right}
A^{q\xi v \pm}_{\alpha\alpha^\prime} = A^{q\xi v \pm}_{{\rm c}}  \,\,\delta_{k\pm q, k^\prime} \,  \, \delta_{b b^\prime}\, \delta_{b {\rm c}}\quad,
\end{equation}
with
\begin{equation}
A^{q\xi v \pm}_{{\rm c}} = \sqrt{
2 \pi \left(N^\circ_{q\xi}+{1 \over 2} \pm {1 \over 2}\right)
\over 
\hbar
}  \, g^{q\xi v}_{{\rm c}}\, D^{q\xi}_{{\rm c}}\quad.
\end{equation}
The related generalized scattering rates in (\ref{calP}) are thus given by
\begin{equation}\label{calP-cond-right}
\mathcal{P}^{\xi v}_{\alpha_1\alpha_2,\alpha^\prime_1\alpha^\prime_2} 
\!=\! \delta_{b_1 b_2, b^\prime_1 b^\prime_2} \delta_{b_1 b_2,{\rm c} {\rm c}} \sum_{q \pm} |A^{q\xi v \pm}_{{\rm c}}|^2 
\delta_{k_1\pm q,k_1^\prime} \delta_{k_2\pm q,k_2^\prime}
\end{equation}
and turn out to be real and positive, just like semiclassical rates (here $\delta_{i_1 i_2, j_1 j_2}$ is a shorthand notation for $\delta_{i_1 j_1} \delta_{i_2 j_2}$).

Inserting Eq.~(\ref{calP-cond-right}) into Eq.~(\ref{calPbar}), 
the explicit form of the linear superoperator (\ref{DME-scat-lin}) corresponding to forward scattering processes acting on right-moving electrons comes out to be
\begin{widetext}
\begin{eqnarray}\label{DME-scat-link}
\left.\frac{d \rho^{v}_{k_1,{\rm c};k_2,{\rm c}}}{d t}\right|_{\rm scat} 
&=& 
\frac{1}{2} \sum_{q\xi \pm} 
\left\{
|A^{q\xi v \pm}_{{\rm c}}|^2 
\left[ (1-f^\circ_{k_1,{\rm c}}) \Delta\rho^{v}_{k_1 \pm q,{\rm c};k_2\pm q,{\rm c}} +  f^\circ_{k_1,{\rm c}} \Delta\rho^{v}_{k_1 \mp q,{\rm c};k_2\mp q,{\rm c}}\right] 
\,+\,{\rm H.c.}\right\}
\nonumber \\
&-& 
\frac{1}{2} \sum_{q\xi \pm} 
\left\{
|A^{q\xi v \pm}_{{\rm c}}|^2 
\left[
\Delta\rho^{v}_{k_1,{\rm c};k_2,{\rm c}} (1-f^\circ_{k_1 \mp q,{\rm c}} + f^\circ_{k_1 \pm q,{\rm c}}) \right]
\,+\,{\rm H.c.}\right\}\ .
\end{eqnarray} \, .
\end{widetext}

The spatial diffusion of the excitation charge density can now be determined by inserting   Eq.~(\ref{DME-scat-link}) into Eq.~(\ref{DME-Deltan_scat}). 
There, by means of a proper rescaling of the sum variables $k_1$ and $k_2$, it is easy to show that the contributions arising from the in- and out-scattering terms (first and second line in (\ref{DME-scat-link})) cancel out, and as a result, one obtains that 
$
d\Delta n^{r}_{\rm c}/dt |_{\rm scat} 
$ vanishes. 
Along quite similar lines of reasoning one can conclude that the same result holds for intraband scattering between left-movers ($\lambda={\rm l}$). 

Finally, we stress that the vanishing effect of forward processes on the spatial diffusion, ascribed to a $k$-space cancellation between in- and out-scattering terms, applies to intraband scattering only.\\


%

\end{document}